\documentclass{elsart}
\usepackage{amsmath}
\usepackage{graphicx}
\usepackage{graphics}
\usepackage{amsfonts}
\usepackage{amssymb}
\journal{Physica D}

\begin{document}
%
\begin{frontmatter}%

\title
{Phase Transitions in Random Boolean Networks with Different Updating Schemes}%

\author{Carlos Gershenson}%

\address
{Centrum Leo Apostel, Vrije Universiteit Brussel. Krijgskundestraat 33. B-1160 Brussel - Belgium}%

\begin{abstract}%
In this paper we study the phase transitions of different types of Random
Boolean networks. These differ in their updating scheme: synchronous,
semi-synchronous, or asynchronous, and deterministic or non-deterministic. It
has been shown that the statistical properties of Random Boolean networks
change considerable according to the updating scheme. We study with computer
simulations sensitivity to initial conditions as a measure of order/chaos. We
find that independently of their updating scheme, all network types have very
similar phase transitions, namely when the average number of connections of
nodes is between one and three. This critical value depends more on the size
of the network than on the updating scheme.%
\end{abstract}%


\end{frontmatter}%

\section*{Introduction}

Random Boolean networks (RBNs) have been used to model several complex
systems. Originally proposed by Kauffman to model genetic regulatory networks
\cite{Kauffman1969}\cite{Kauffman1993}, their study has propagated into
different research areas, from mathematics to robotics, from biology to
artificial life. It has been noticed that the properties and behaviour of
RBNs, such as average number of attractors and shape of attractor basins,
change considerably depending on the updating scheme we use
\cite{HarveyBossomaier1997}\cite{Gershenson2002e}\cite{GershensonEtAl2003a}.
For this paper, we studied the following question:\ Does the updating scheme
affect the dynamical phase (ordered, chaotic) of the network? In other words,
for a given family of RBNs, if we change the updating scheme, could this also
change the regime?

In the following section, we review the properties of different types of RBNs
according to their updating scheme: synchronous or asynchronous, deterministic
or non-deterministic \cite{Gershenson2002e}. In Section \ref{sect-phasetrans}
we mention previous studies on phase transitions of RBNs. In Section
\ref{sect-experiments} we present experiments we performed for comparing the
sensitivity to initial conditions of different types of RBN. The results are
discussed in Section \ref{sect-discussion}, for concluding in Section
\ref{sect-conclusions}.

\section{Different Types of Random Boolean Networks}

\bigskip Random Boolean networks consist of $n$ nodes or units which can take
values of zero or one. The actual value of each node is controlled by the
values of $k$ nodes which are ``connected'' to it, and determined by lookup
tables of logic functions. The lookup tables and the connectivity of the
network are generated randomly, but remain fixed through the dynamics of the
network. RBNs are interesting models because one does not assume the
functionality of the network in order to study the generic properties of a
type of networks. The parameters which affect the behaviour of networks
include number of nodes, number of connections, topology
\cite{OosawaSavageau2002}\cite{Aldana2003}, and updating scheme
\cite{HarveyBossomaier1997}\cite{Gershenson2002e}\cite{GershensonEtAl2003a}.

\textbf{Classical Random Boolean networks} (CRBNs) were proposed by Kauffman
to model genetic regulatory networks \cite{Kauffman1969}\cite{Kauffman1993}.
He used synchronous updating:\ the values of the nodes at time $t+1$ depend of
the values of the nodes at time $t$. Since they are deterministic, and have a
finite number of nodes, sooner or later they reach a state which was visited
already. The network has fallen into an attractor. If the attractor consists
of a single state, it will be called a point attractor, whereas a cycle
attractor consists of several states. CRBNs have been widely studied
(\textit{e.g.} \cite{Wuensche1997}\cite{AldanaEtAl2003})

\textbf{Asynchronous Random Boolean networks} (ARBNs) were developed by Harvey
and Bossomaier \cite{HarveyBossomaier1997}, criticizing the assumption of
Kauffman that genetic regulatory networks are synchronous. ARBNs take one node
at random each time step and update it. The behaviour of the network changes
considerably from CRBNs. The basins of attraction change, and cycle attractors
disappear, since these networks are non-deterministic in their updating, and
the sequence of transitions will not be repeated. However, DiPaolo was able to
use genetic algorithms to find ARBNs with\ ``rhythmic'' attractors
\cite{DiPaolo2001}. Also, it has been shown that asynchronous dynamical
systems can reproduce the behaviour of synchronous systems\cite{Nehaniv2002}%
\cite{KlemmBornholdt2003}.

\textbf{Deterministic Asynchronous Random Boolean networks }(DARBNs)
\cite{Gershenson2002e} were defined criticizing non-determinism in ARBNs, and
as an intermediate alternative between CRBNs and ARBNs. Each node can be
updated in different time periods, but these are deterministic (although
randomly generated). We need for each node parameters $p$ and $q$ which
determine the ``period'' and the ``translation'' of the update. So each node
is updated when the modulus of time over $p$ equals $q,$ ($p>q$). If more than
one node fulfills this condition, the network is updated in a specific
sequence for each node, one by one, to simulate the updating of ARBNs. Since
the functions, topology, and update parameters are generated randomly, the
sequence does not affect the general properties of a family of networks.
DARBNs have again cycle attractors, and their properties are similar to the
ones of CRBNs. However, as we allow a wider variety of updating periods
(increasing $maxP$, the maximum $p$ allowed), the properties are more similar
to the ones of ARBNs.

\textbf{Generalized Asynchronous Random Boolean networks }(GARBNs)
\cite{Gershenson2002e} are very similar to ARBNs, only that they choose
randomly at each time step which nodes to update, and these are updated
synchronously. Therefore, there could be none, one, some, or all of the nodes
being updated at a particular time step. We can say that GARBNs are semi-synchronous.

\textbf{Deterministic Generalized Asynchronous Random Boolean networks
}(DGARBNs) \cite{Gershenson2002e}, like DARBNs also use parameters $p$ and $q$
to make the dynamics deterministic. The difference with DARBNs is that when
several nodes fulfill the updating condition, all them are updated
synchronously. Thus, DGARBNs have semi-synchronous but deterministic updating.
CRBNs are a particular case of DGARBNs, ($maxP=1$). We can describe the set of
parameters $p$ and $q$ as the \emph{context} of the network. We have used
DGARBNs as models of contextual discrete dynamical systems with promising
results \cite{GershensonEtAl2003a}. We have found that changing the context of
a particular network can change considerably its attractor basins and number
of attractors, without changing the rules or connectivity of the network.

Changing the updating scheme of a RBN changes its behaviour considerably. Even
among deterministic RBNs, the attractor basins shift for different updating
schemes. Different types of RBNs have different statistical properties, such
as average number of attractors and percentage of states in attractors.
However, we have realized that point attractors are the same for all types of
RBNs \cite{Gershenson2002e}, since no matter which node is updated, the state
will not change. Thus, the updating does not affect the point attractors. Also
all the RBNs have very similar behaviour for the anomalous case $k=0$, because
there are no actual dynamics: all initial states tend to a single point attractor.

We have shown \cite{GershensonEtAl2003a} that there is a smooth transition of
the statistical properties of RBNs from deterministic to asynchronous
non-deterministic, passing through asynchronous deterministic as we allow a
wider variety of their updating periods (\textit{i.e.} increase $maxP$).

\section{Phase Transitions in RBNs\label{sect-phasetrans}}

\bigskip Random Boolean networks can have different dynamic regimes. On one
hand, we have dynamics which stabilize quickly, and are robust to
perturbations, which are called ordered. On the other hand, we have a chaotic
regime, where networks take time to reach an attractor (this can be infinite
in practice), and are very sensitive to initial conditions. Kauffman
conjectured that biological networks should be ``at the edge of chaos'', with
enough flexibility to explore new functionalities, but enough robustness to
keep the present ones.

\bigskip Classical RBNs with uniform connectivity have been found to have a
phase transition from ordered to chaotic for a critical value for $k=2$
\cite{DerridaPomeau1986}. Bias in the functions (not an equal opportunity for
zeros and ones) and topology of the networks have been shown to change the
critical value \cite{BallesterosLuque2002}\cite{Aldana2003}

\bigskip Mesot and Teuscher \cite{MesotTeuscher2003} studied critical values
for GARBNs inspired in the annealed approximation method
\cite{DerridaPomeau1986}, but found no critical value, in the sense that
GARBNs amplify small perturbations and reduce big ones. The results we
obtained measuring sensitivity to initial conditions question their findings.

\section{Experiments\label{sect-experiments}}

\bigskip We carried out experiments in an open source software laboratory,
RBNLab, which we developed for studying different properties of Random Boolean
networks. It is available to the public (Java source code included)
\cite{RBNLab}.

For our experiments, we used ``standard'' RBNs: homogeneous probability in the
boolean functions (no bias for more zeroes or ones) and uniform topology (all
nodes have exactly $k$ connections).

There are several features which can be used to measure chaos, and which can
yield different results. We decided to take sensitivity to initial conditions
as a measure for chaos, which is similar to the study of minimal perturbations
or damages \cite{BallesterosLuque2002}. We first created randomly an initial
state A, and flip one node to have another initial state B. We run each
initial state in the network for ten thousand time steps, obtaining states A'
and B'. Then we compare the normalized Hamming distance (\ref{hamming}) of the
final states with the one of the initial states to obtain a parameter $\delta$
(\ref{delta}).%

\begin{equation}
H(A,B)=\frac{1}{n}\sum\limits_{i}^{n}\left\vert a_{i}-b_{i}\right\vert
\label{hamming}%
\end{equation}%

\begin{equation}
\bigskip\delta=H_{t\rightarrow\infty}-H_{t=0} \label{delta}%
\end{equation}

If $\delta$ is negative, it means that the Hamming distance was reduced. Since
the initial distance is minimal ($\frac{1}{n}$), a negative $\delta$ indicates
that both initial states tend to the same attractor. This implies that the
network is stable, or on an ordered phase. A positive $\delta$ indicates that
the dynamics for very similar initial states diverge. This is a common
characteristic of chaos in dynamical systems.

Since the initial states are chosen randomly, the comparison we make is
equivalent to see B as a perturbed version of A, and observe if the
perturbation affects the dynamics.

To compare the regimes of different types of RBN, we created $NN$ number of
networks (200 unless specified), and evaluated for each $NS$ number of states
(200 unless specified) for all five types of RBN (using $maxP=7$ for the
deterministic asynchronous RBNs), for different network sizes ($n$) and
connectivity density ($k$).

\begin{figure}
[ptb]
\begin{center}
\includegraphics[
natheight=6.167000in,
natwidth=9.979100in,
height=3.1107in,
width=5.0176in
]%
{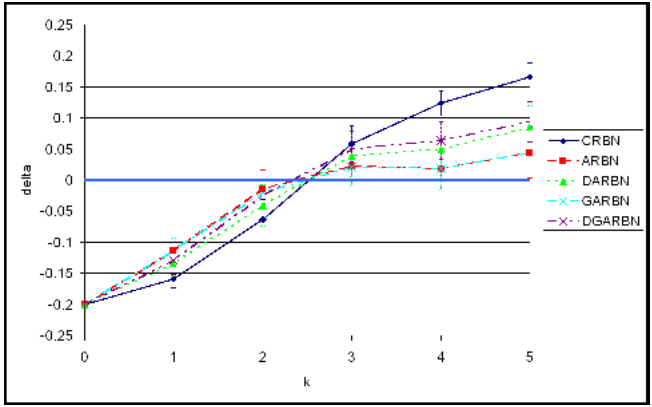}%
\caption{Sensitivity to initial conditions for networks with $n=5$}%
\label{n=5}%
\end{center}
\end{figure}

We can observe the averages of $\delta$\ for networks with $n=5$ in Figure
\ref{n=5}. The error bars indicate the standard deviations. We can see that
all networks have an average phase transition from ordered to chaotic for
values of $k$ between two and three (although the standard deviations indicate
us that there can very well be chaotic networks for $k=2$ and ordered for
$k>2$). It is curious to note that for this network size, the ``most ordered''
updating schemes of the ordered phase are also the ``most chaotic'' of the
chaotic phase, and vice versa.%

\bigskip Independently of network size, all RBNs will have $\delta=-\frac
{1}{n}$ for $k=0$, since all states tend to a single point attractor.

\begin{figure}
[ptb]
\begin{center}
\includegraphics[
natheight=6.167000in,
natwidth=9.979100in,
height=3.1116in,
width=5.0168in
]%
{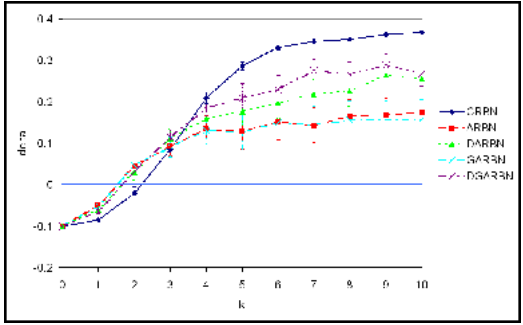}%
\caption{Sensitivity to initial conditions for networks with $n=10$}%
\label{n=10}%
\end{center}
\end{figure}

\bigskip In Figure \ref{n=10}, we can observe that for $n=10$ all network
types except CRBN are already on the chaotic regime for $k=2$, although still
close to the ``edge of chaos''.%

\begin{figure}
[ptb]
\begin{center}
\includegraphics[
natheight=6.167000in,
natwidth=9.979100in,
height=3.1116in,
width=5.0168in
]%
{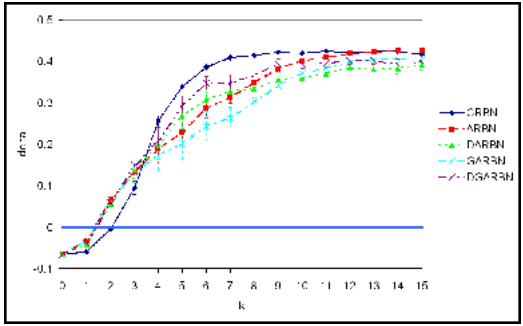}%
\caption{Sensitivity to initial conditions for networks with $n=15$}%
\label{n=15}%
\end{center}
\end{figure}

\bigskip\bigskip This is similar for $n=15$, although CRBN has average
$\delta$ values very close to zero, as we can see in Figure \ref{n=15}.%

\begin{figure}
[ptb]
\begin{center}
\includegraphics[
natheight=6.167000in,
natwidth=9.979100in,
height=3.1116in,
width=5.0168in
]%
{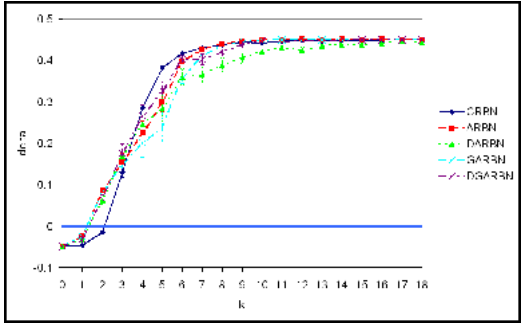}%
\caption{Sensitivity to initial conditions for networks with $n=20$}%
\label{n=20}%
\end{center}
\end{figure}

As we increase $n$, the point where $\delta$ values of CRBNs cross to be
higher than the others shifts towards larger values of $k$. As shown in Figure
\ref{n=20}, for $n=20$ and large values of $k$, DARBNs turn out to be ``less
chaotic'' than the asynchronous non-deterministic networks. For large values
of $k$ and $n$, all networks tend asymptotically to an average maximum value
of $\delta$, where almost all close states tend to different attractors,
independently on the updating scheme.%

\begin{figure}
[ptb]
\begin{center}
\includegraphics[
natheight=6.167000in,
natwidth=9.979100in,
height=3.1116in,
width=5.0168in
]%
{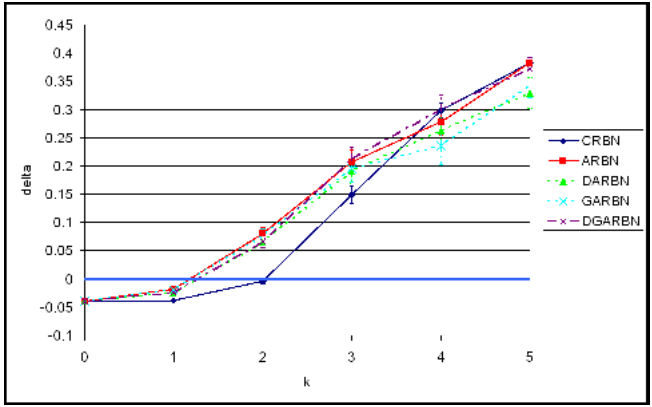}%
\caption{Sensitivity to initial conditions for networks with $n=25$}%
\label{n=25}%
\end{center}
\end{figure}

\begin{figure}
[ptb]
\begin{center}
\includegraphics[
natheight=6.167000in,
natwidth=9.979100in,
height=3.1116in,
width=5.0168in
]%
{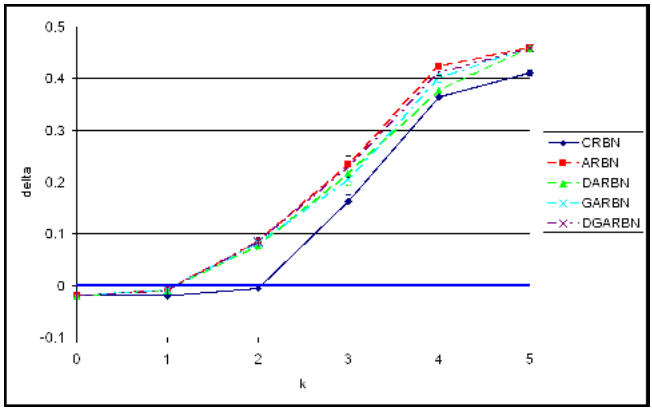}%
\caption{Sensitivity to initial conditions for networks with $n=50$. NN=100,
NS=100}%
\label{n=50}%
\end{center}
\end{figure}

\begin{figure}
[ptb]
\begin{center}
\includegraphics[
natheight=6.167000in,
natwidth=9.979100in,
height=3.1116in,
width=5.0168in
]%
{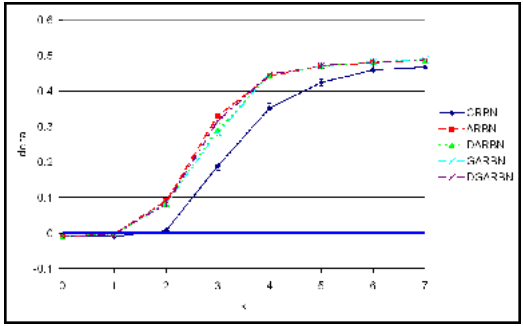}%
\caption{Sensitivity to initial conditions for networks with $n=100$. NN=50,
NS=50}%
\label{n=100}%
\end{center}
\end{figure}

\begin{figure}
[ptb]
\begin{center}
\includegraphics[
natheight=6.167000in,
natwidth=9.979100in,
height=3.1116in,
width=5.0168in
]%
{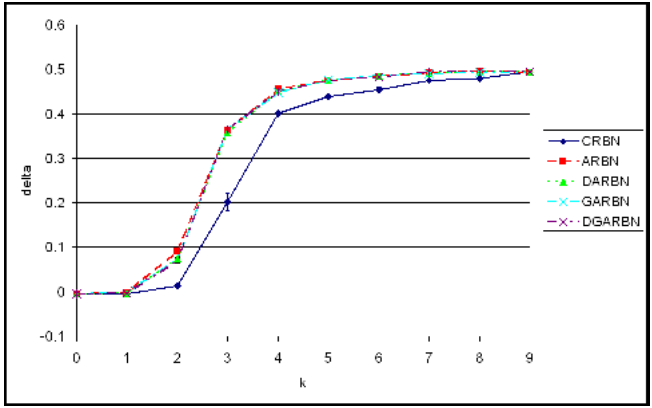}%
\caption{Sensitivity to initial conditions for networks with $n=200$. NN=20,
NS=20}%
\label{n=200}%
\end{center}
\end{figure}

\bigskip A similar pattern follows as we increase the network size, as seen in
Figures \ref{n=25}, \ref{n=50}, \ref{n=100}, and \ref{n=200}. What changes is
that for very large networks, CRBNs seem to be always ``less chaotic'' than
others, even in the chaotic phase.%

\bigskip It is interesting to note that the standard deviation decreases as we
increase the size of the networks. This is good, since larger networks need
more computational resources to be analysed. We should also note that always
the highest variance is around $k=3$. This indicates a wider diversity in the
types of networks. This is a desirable property in evolving networks. It
should be mentioned that also at k=3 point attractors are ``harder to find''
(they have the most skewed distribution among all possible networks: few
networks with lots of point attractors, many with none)
\cite{HarveyBossomaier1997}.%

\bigskip We can see that for very large networks ($n>100$), even CRBNs with
$k=2$ fall into the chaotic regime.%

The detailed data and graphics obtained for these experiments can be found
online \cite{vub.rbn}.

\section{Discussion\label{sect-discussion}}

\bigskip Even when there are differences in the average values of $\delta$ for
different types of RBN, all of them follow a similar path, crossing from
ordered to chaotic for values $1<k<3$. The precise value of the transition, in
practice, does not depend only on the updating scheme, but also on the size of
the network. However, we should note that our results show only averages, and
that it is always possible to find ordered RBNs of an arbitrary size and
connectivity\footnote{An example can be a network which states represent
numbers in base two. If each state goes to the state representing the next
ascending number, and the last (2$^{n}$) to itself, we have a very ordered
network independently of $n$.}. What our results show is that networks with
complex behaviour are easier to find for values $1<k<3$, independently of the
updating scheme. This is consistent with previous results for CRBNs. It is
interesting that in practice smaller networks have transitions at values of
$k$ closer to three, while larger networks have values closer to one.

Mesot and Teuscher \cite{MesotTeuscher2003} did not find any phase transition
for GARBNs. They used a different approach, namely the annealed approximation
method\cite{DerridaPomeau1986}. In their simulations, they studied networks of
$n=200$ and $k>=1$. As we can see in Figure \ref{n=200}, the behaviour for
$k=1$ is very similar than the one for $k=0$. It seems that the explanation
for this is that very large networks with only one connection are ``frozen''
easily, because with an uniform topological distribution, it is hard for
perturbations to propagate. On the other hand, if we have two connections per
node, distant nodes can affect each other indirectly. This explains why larger
networks tend to be less stable, even with the same number of connections. For
$n=200$, the behaviour of the networks is on average either static ($k<=1$) or
chaotic ($k>=2$), but \emph{independently} of the updating scheme. It seems
that the edge of chaos ``shrinks'' as we increase the size of networks. In
other words, complex behaviour will still be found somewhere near a certain
critical value, only that on average it will be harder to find particular
networks which exhibit complex behaviour, i.e. for larger networks there is a
lower percentage of complex networks in all possible networks.

It is also worth noting that for large networks, asynchronous updating
schemes, deterministic or not, tend to be very similar in terms of $\delta$.
It seems that synchronicity plays a more important role in phase transitions
than the determinism of the updates. This is opposite to the network
properties, where the determinism makes a greater difference than the
synchronicity \cite{Gershenson2002e}, mainly because cycle attractors are lost
in non-deterministic updating schemes. Deterministic asynchronous RBNs have
the advantage that they are easier to study than the non-deterministic ones,
but we have seen that they have similar phase transitions. This is yet another
reason for preferring GARBNs for modelling complex
networks\cite{GershensonEtAl2003a}, although specific considerations can
suggest another network type.

GARBNs could also be preferred because they can ``process more information''
than other types of networks of the same size and connectivity, since they can
``throw information into their context''\cite{GershensonEtAl2003a}. This is
because any deterministic asynchronous RBN can be mapped into a redundant
CRBN, incorporating the information of the updating periods into the network
\cite{Gershenson2002e}. But this takes the RBN into a family of networks of
the chaotic phase, since we add more connections. With the results presented
in this paper, we can see that if we vary the updating periods of a RBN, the
phase would still be similar. This allows the same network with the same rules
and topology to have different functionality in different situations or
contexts. Therefore, more information can be processed by DGARBNs by
complexifying its updating periods, but without moving the dynamics into the
chaotic phase. This is an evolutionary advantage, and we should expect natural
systems to exploit their context or environment in order to process more
information without increasing their own information processing capacity. An
example of this are humans using their environment (books, computers, culture)
to enhance their cognitive abilities\cite{ClarkChalmers1998}.

The role of the topology on the properties and behaviour of RBNs should be
studied for different types of RBNs, especially for scale-free topologies,
following the work of Aldana \cite{Aldana2003}.

\section{\bigskip Conclusions\label{sect-conclusions}}

\bigskip We have tested experimentally the sensitivity to initial conditions
of different types of Random Boolean networks. We found no significant
difference between the different updating schemes and the phase transition.
All RBNs have a phase transition from ordered to chaotic for values $1<k<3$,
independently of their updating scheme. In practice, the precise critical
value of $k$ depends more on the size of the network than on the updating
scheme. This is interesting because the statistical properties, such as
average number of attractors and average attractor length, do change with the
updating scheme. It implies that the phase of a RBN (ordered, critical,
chaotic) does not depend really on the updating scheme, but on the network
size, connectivity (and topology), and functionality of the network.

\section{Acknowledgements}

I would like to thank Diederik Aerts, Francis Heylighen, Jan Broekaert, Ricard
Sol\'{e}, and Pau Fern\'{a}ndez for interesting discussions and suggestions.
This work was funded in part by the Consejo Nacional de Ciencia y
Tecnolog\'{i}a (CONACyT) of Mexico.

\bibliographystyle{elsart-num}
\bibliography{carlos,COG,RBN}
\end{document}